# Repurposing drugs for COVID-19 based on transcriptional response of host cells to SARS-CoV-2

Fuhai Li[1,2,#], Andrew P. Michelson[1,3], Randi Foraker[1], Ming Zhan[4], Philip R.O. Payne[1]

[1]Institute for Informatics (I2), [2]Department of Pediatrics, [3]Pulmonary and Critical Care Medicine, Washington University in St. Louis School of Medicine, St. Louis, MO, USA; [4]National Institute of Mental Health (NIMH), NIH, Bethesda, MD, USA.

[#]Correspondence Email: Fuhai.Li@wustl.edu

**Abstract**

The Coronavirus Disease 2019 (COVID-19) pandemic has infected over 10 million people globally with a relatively high mortality rate. There are many therapeutics undergoing clinical trials, but there is no effective vaccine or therapy for treatment thus far. After affected by the Severe Acute Respiratory Syndrome Coronavirus 2 (SARS-CoV-2), molecular signaling of host cells play critical roles during the life cycle of SARS-CoV-2, like RNA replication, assembly. Thus, it is significant to identify the involved molecular signaling pathways within the host cells, and drugs targeting on these molecular signaling pathways could be potentially effective for COVID-19 treatment. In this study, we aimed to identify these potential molecular signaling pathways, and repurpose existing drugs as potential effective treatment of COVID-19 to facilitate the therapeutic discovery, based on the transcriptional response of host cells. We first identified dysfunctional signaling pathways associated with the infection caused SARS-CoV-2 in human lung epithelial cells through analysis of the altered gene expression profiles. In addition to the signaling pathway analysis, the activated gene ontologies (GOs) and super gene ontologies were identified. Signaling pathways and GOs such as MAPK, JNK, STAT, ERK, JAK-STAT, IRF7-NFkB signaling, and MYD88/CXCR6 immune signaling were particularly identified. Based on the identified signaling pathways and GOs, a set of potentially effective drugs were repurposed by integrating the drug-target and reverse gene expression data resources. In addition to many drugs being tested in clinical trials, the dexamethasone was ranked top 1 in our analysis, which was the first reported drug to be able to significantly reduce the death rate of COVID-19 patients receiving respiratory support. These results can be helpful to understand the associated molecular signaling pathways within host cells, and facilitate the discovery of effective drugs for COVID-19 treatment.

# 1. Introduction

By June 29, 2020, there were over 2,500,000 confirmed cases (with >120,000 deaths) of *Coronavirus Disease 2019 (*COVID-19) in the U.S. and over 10 million cases (with >500,000 deaths) globally, based on the COVID-19 Dashboard[1] operated by the Center for Systems Science and Engineering (CSSE) at Johns Hopkins University (JHU) (https://coronavirus.jhu.edu/map.html). The primary organ of infection is considered to be the lung, and the infection leads to acute hypoxemic respiration and ultimately to multi-organ failure and death[2]. The mortality rate of COVID-19 is relatively high[3], compared with the flu epidemic. So far, there is no newly FDA approved drug for the treatment of COVID-19. Recently, *remdesivir*, developed by Gilead Sciences, was granted an FDA emergency use authorization for COVID-19 treatment. However, *remdesivir* can reduce the time of recovery and cannot significantly reduce the mortality rate[4]. To improve the outcome of COVID-19 patients, many existing drugs are being evaluated in clinical trials globally, like *chloroquine and hydroxychloroquine, azithromycin, and lopinavir–ritonavir, and* dexamethasone*.* Thus, repurposing existing medications is considered an important approach to speed up the drug discovery for COVID-19. A few days ago, dexamethasone, an existing FDA approved drug, was reported to be the first drug that can reduce the death rate, by one-fifth to one-third, of COVID-19 patients receiving respiratory support[5], which was ranked No.1 in our analysis.

Although 1,570 clinical trials have been initiated globally for COVID-19 treatment by June 29, 2020, based on the data from the dashboard[6] of real-time clinical trials of COVID-19 (https://www.covid-trials.org/), only one drug, dexamethasone, was reported to be able to significantly reduce the death rate of COVID-19 patients receiving respiratory support[5]. One possible reason is that most of the current clinical trials are based on limited knowledge of the disease and observed phenotypes. The molecular mechanisms and signaling pathways within the host cells such as lung cells, which play critical roles in the life cycle of *Severe Acute Respiratory Syndrome Coronavirus 2 (*SARS-CoV-2) infection, remain unidentified. Thus, it is significant to uncover the mysterious molecular signaling pathways within host cells via computational data analysis. It is also important and needed to facilitate drug repurposing and design of new clinical trials.

In order to understand the transcriptional response of lung cells to the SARS-CoV-2 infection, Albrecht and tenOever laboratories profiled the RNA-seq gene expression from human NHBE (Normal Human Bronchial Epithelial) cells, A549 lung cancer cells (no ACE2 expression), A549_ACE2 (A549 lung cancer cells transduced with a vector expressing human ACE2), and CALU-3 lung cancer cells (with ACE2 expression), and 2 human lung samples infected by SARS-

CoV-2[7]. The data are valuable sources for identifying genetic pathways that become dysregulated during active infection. More importantly, the data allows for the identification of activated signaling pathways that can be targeted by existing pharmaceutical agents. In this study, we aimed to identify activated signaling pathways within lung host cells affected by SARS-CoV-2 and repurpose existing drugs for COVID-19 treatment using a novel integrative data analysis approach, integrating transcriptional response[7], signaling pathway[8], gene ontology[9], drug-target interactions from drugbank[10] and reverse gene signature data from connectivity map (CMAP)[11,12]. These results, including the identified signaling pathways, activated GOs, and drugs, can be helpful to facilitate the experimental screening and clinical trial design to speed up the therapeutic discovery for COVID-19.

## 2. Materials and Methodology

RNA-seq data (gene expression) from NHBE (Normal Human Bronchial Epithelial) cells, A549 (no ACE2 expression) cells, A549_ACE2 cells (A549 lung cancer cells transduced with a vector expressing human ACE2), and CALU-3 lung cancer cells (with ACE2 expression) cells infected by SARS-CoV-2 were obtained from GEO (GSE147507)[7]. This data was generated by Drs. Albrecht and tenOever's at the Icahn School of Medicine at Mount Sinai. The DEseq2[13] tool was used to calculate the fold change and p-value of individual genes in the NHBE (normal tissue), A549_ACE2, and CALU-3 lung cancer cells respectively before- and after- viral exposure. The data of A549 cell was not used considering that the ACE2, with which SARS-CoV-2 interacts to enter host cells, is not expressed in A549 cell.

For the signaling network analysis, the 307 KEGG (Kyoto Encyclopedia of Genes and Genomes)[8] signaling pathways were extracted. To identify the activated signaling network for NHBE, A549_ACE2 and CALU-3 cells respectively, the signaling paths, i.e., the shortest paths link source signaling genes (starting genes on the signaling pathways) and the sink signaling genes (ending genes on the signaling pathways) within the 307 KEGG signaling pathways were first identified. For each signaling path, the average fold change of genes on the signaling path was calculated. Then the signaling paths with the average fold change score greater than the mean score + 1.25 standard deviation threshold were selected to construct the activated signaling network. To remove the tumor specific signaling pathways, the common signaling pathways between NHBE and A549_ACE2, and between NHBE and CALU-3 were unified as the potential activated signaling pathways associated with the viral infection. To identify potential drugs that can inhibit the signaling genes on the activated signaling pathways, the drug-target interactions of FDA-approved drugs were downloaded from the DrugBank[10] database. Then drugs targeting the activated signaling pathways were identified as potential effective drug candidates for COVID-19 treatment.

For the gene ontology (GO)[9] analysis, the Fisher's exact test, with a threshold p-value = 0.1, was used to identify the statistically activated GOs based on the up-regulated genes. Since there are many activated GOs, and some of them are semantically close and sharing the common set of genes, it was difficult to identify the most important GOs. To solve this challenge, we first manually removed many of the activated GOs that were not related to biological signaling processes or general diseases. In addition, we defined the super-GOs, which are defined as sub-groups of GOs that have similar or related biological processes. Specifically, after the calculation of GO-GO similarity using the semantic similarity[14] (GOSemSim R package) between activated GOs, the affinity propagation clustering[15] (APclustering) was used to divide the activated GOs into sub-groups (named super-GOs). Then, the genes that were up-regulated within each super-GO were used as signatures to identify potential drugs that can inhibit activation of the super-GOs. These gene signatures were fed into the connectivity map (CMAP)[11,12] database to identify potential drugs. A gene set enrichment analysis (GSEA)[16] was applied on the z-profiles (gene expression variation before and after treatment with 2,837 drugs and investigational agents) of 9 cells[11,12] in the CMAP to identify gene set signature-specific inhibitory drugs. The top ranked FDA drugs, based on the average GSEA scores, that can potentially inhibit the up-regulated gene signatures associated with the super-GOs were identified as potential candidates for COVID-19 treatment.

## 3. Results

*3.1 Activated signaling pathways and associated inhibitory medications*

The KEGG signaling pathway analysis was conducted to identify the activated signaling pathways within the lung A549 cells after SARS-CoV-2 infection (see **Fig. 1-Upper**). As shown, the IRF7/IFR9, NFkB1, NFkB2, STAT1, TNF, MAPK3K8, MAPK8, and MAPK14 related signaling pathways are the major activated transcription factors (TFs), which promotes the downstream activation of many signaling pathways including those mediated by CXCR6/CXCL1/CXCL2/CXCL3/CXCL10, MYD88, CREBBP, JAK1/JAK2, STAT and MAPK signaling pathways. Also, the WNT4/WNT7A and SMAD signaling pathways are also activated.

Moreover, the PDGFB-EGFR and TUBB1C/2B/3 proliferation signaling pathways are also activated.

Based on the drug-target interaction data derived from DrugBank, there are 220 drugstargeting 97 genes on the signaling network, with 71 drugs targeting on the PTGS2 gene specifically (see **Table S1**). Chloroquine and hydroxychloroquine were found to inhibit MYD88 immune signaling. Also, acetylsalicylic acid, thalidomide, pranlukast, triflusal, glycyrrhizic acid and fish oil were found to inhibit NFkB signaling pathway, which was a potential signaling target for SARS[17] treatment, and can also potentially inhibit IRF7 activity. Thalidomide[18] inhibiting NFkB and TNF was reported as potential treatment for COVID-19. Moreover, the tumor necrosis factor (TNF) found in this study is reported in Lancet[19] to be an important therapeutic target for COVID-19. Also, JAK1/2 pathways were reported as important targets, and their inhibitors, ruxolitinib, tofacitinib, baricitinib and fostamatinib, could be effective for COVID-19 treatment. Particularly, baricitinib, an arthritis drug, could help reduce the out-of-control immune response (https://www.wired.com/story/ai-uncovers-potential-treatment-covid-19-patients/, and https://www.clinicaltrialsarena.com/news/eli-lilly-to-study-baricitinib-for-covid-19-treatment/), was reported in the Lancet[20] as a COVID-19 suitable treatment. In addition, the MAPK1, AKT and PRKCA inhibitors such as isoprenaline, arsenic trioxide, vitamin e, and midostaurin could be also effective. Moreover, the IL6R inhibitor tocilizumab was reported for COVID-19 treatment[21], and another IL6R inhibitor, sarilumab, is being evaluated in clinical trials (ClinicalTrials.gov Identifier: NCT04327388). Lastly, STAT1 and IFANR1 were identified as potential targets for COVID-19 treatment and were reported by another group as well[22]. These results indicated that the identified activated signaling pathways might play important roles in the viral life cycle and can be helpful to identify potential therapeutic targets and drugs.

In addition, drugs that have been tested or are currently being tested in clinical trials globally were identified from the covid-trials dashboard (URL: https://www.covid-trials.org/ (Data: table_trials - 2020-05-27 05_38_12.csv, and the updated drugs: table_trials - 2020-06-29 22_20_46.csv. Some drugs cannot be found in the table_trials - 2020-06-29 22_20_46.csv because drug names were replaced by using category names, e.g., *ruxolitinib was replaced by* JAK inhibitor), and compared with the drugs identified from the signaling pathway analysis. Based on the dashboard of clinical trials for COVID-19 treatment, ~114 FDA approved drugs were reported from 1,132 clinical trials globally (see **Table I**). Among the 108 drugs, 31 of them are in the predicted drugs list. These drugs are: *hydroxychloroquine, chloroquine, tocilizumab, sarilumab, canakinumab, ruxolitinib, oxygen, sirolimus, ibuprofen, anakinra, acalabrutinib, baricitinib, ibrutinib, lenalidomide, tirofiban, acetylsalicylicacid, simvastatin, siltuximab,*

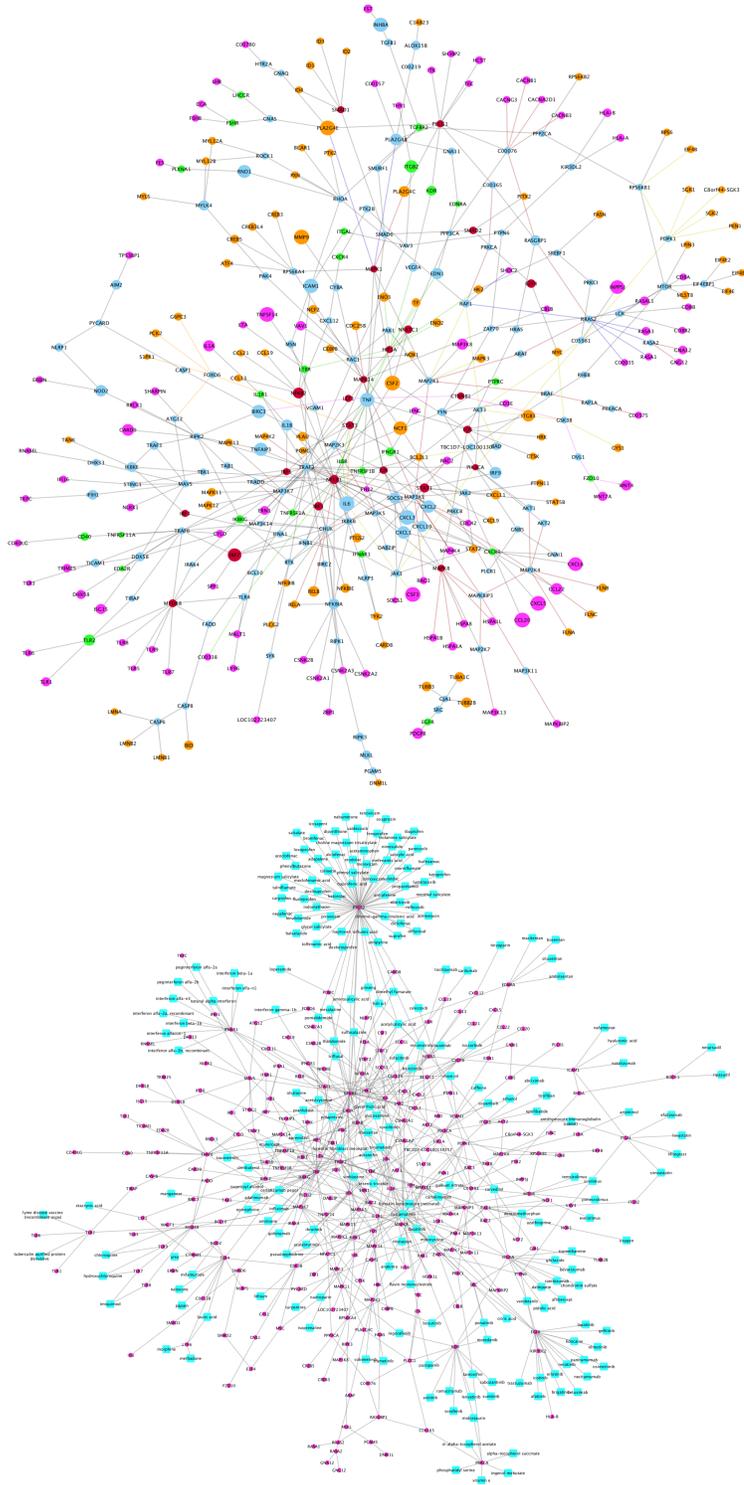

**Figure 1**: (**Upper**) Activated signaling pathways (with 244 genes) within NHBE and also appeared in lung A549_ACE2 and CALU3 cells after SARS-CoV-2 infection. Red nodes are transcription factors (TFs); green - receptors; purple - ligands; orange - activated target genes of TFs; and cyan - the linking genes. (**Lower**) 220 drugs (cyan) (with 97 target genes) targeting on the activated signaling pathways.

*bevacizumab, tinzaparin, naproxen, celecoxib, tofacitinib, adalimumab, thalidomide, dalteparin, nadroparin, minocycline, lithium, indomethacin, and nintedanib*. In summary, the predicted signaling network could be helpful to understand the molecular mechanisms within lung host cells after SARS-CoV-2 infection. Drugs targets on the signaling targets might be effective (some might also have harmful effects and cautions). Drug combinations (drug cocktails) targeting on different targets can be potentially synergistic for COVID-19 treatment.

**Table I**: FDA approved drugs in clinical trials for COVID-19 treatment.

| | | | | |
|---|---|---|---|---|
| zidovudine | linagliptin | chlorpromazine | bevacizumab | sofosbuvir |
| hydroxychloroquine | telmisartan | lenalidomide | tinzaparin | vitamina |
| lopinavir | anakinra | methotrexate | naproxen | metformin |
| tocilizumab | vitaminc | tirofiban | ritonavir | berberine |
| sarilumab | zinc | clopidogrel | tacrolimus | licorice |
| atazanavir | almitrine | acetylsalicylicacid | celecoxib | bromhexine |
| tranexamicacid | sitagliptin | fondaparinux | tofacitinib | minocycline |
| alteplase | ciclesonide | ramipril | pirfenidone | lithium |
| canakinumab | acalabrutinib | progesterone | hydrogenperoxide | formoterol |
| ruxolitinib | etoposide | captopril | sildenafil | indomethacin |
| colchicine | ketamine | eculizumab | ixekizumab | selenium |
| leflunomide | losartan | sevoflurane | dexmedetomidine | nintedanib |
| oxygen | valsartan | nitazoxanide | lipoicacid | spironolactone |
| sirolimus | baricitinib | sargramostim | tranilast | imatinib |
| povidone-iodine | fluoxetine | ribavirin | adalimumab | estradiol |
| fluvoxamine | vitamind | nivolumab | thalidomide | chloroquine |
| ibuprofen | bicalutamide | melatonin | fingolimod | azithromycin |
| aviptadil | ivermectin | simvastatin | suramin | dexamethasone |
| doxycycline | sodiumbicarbonate | dapagliflozin | itraconazole | oseltamivir |
| enoxaparin | ibrutinib | amiodarone | mefloquine | amoxicillin |
| prazosin | levamisole | verapamil | dalteparin | clavulanate |
| isotretinoin | deferoxamine | siltuximab | nadroparin | darunavir |
| heparin | methyleneblue | defibrotide | iloprost | |

*3.2 Activated Gene Ontologies (GOs)*

Based on the fold change and p-value obtained from the DEseq2 analysis of NHBE, A549_ACE2, and CALU-3 cells before and after the viral infection, the up-regulated genes in each cell were identified respectively. Specifically, for the NHBE cells, 558 genes had a statistically significant increase with a fold-change >= 1.25 with a p_value <= 0.05. For the A549_ACE2 cells, 1335 up-regulated genes were identified with a fold-change >= 2.0 and with a p_value <= 0.05. For the CALU-3 cells, 1335 up-regulated genes were identified with a fold-change >= 2.0 and with a p_value <= 0.05.

Based on the three up-regulated gene sets, the activated GOs with enriched genes in the three gene sets were identified respectively with a p-value <= 0.05 (obtained from the Fisher's exact test) and the number of genes in GOs between 10 and 1000. Then the common activated GOs between NHBE and A549_ACE2 cells and between NHBE and CALU-3 cells were unified. After empirically removing the unrelated and general GOs, 73 GOs were kept. Then, the clustering analysis was employed to divide the activated GOs into 5 sub-groups (named super-GOs). Among the 73 GOs, 212 up-regulated genes were kept. **Fig. 2** shows the network of the 212 up-regulated genes, 73 activated GOs and 5 super-GOs. Genes associated with the GOs in each super-GO are collected as gene set signatures for drug discovery analysis. In addition to the viral process related signaling, the GO analysis identified many potential viral infection related signaling pathways, e.g., MAPK, JNK, STAT, ERK1/2, MYD88 and Toll like receptor signaling pathways. These results are consistent and complement the KEGG signaling pathway analysis.

**Table II**: Twenty-seven common FDA approved drugs derived from signaling network analysis and super-GO analysis.

| caffeine | lenalidomide | naproxen | sunitinib | talniflumate |
|---|---|---|---|---|
| lovastatin | dextromethorphan | phenylbutazone | resveratrol | fostamatinib |
| lidocaine | diclofenac | thalidomide | nimesulide | chloroquine |
| gefitinib | chloroquine | naloxone | pazopanib | |
| sorafenib | simvastatin | dasatinib | tofacitinib | |
| nabumetone | imiquimod | lapatinib | afatinib | |

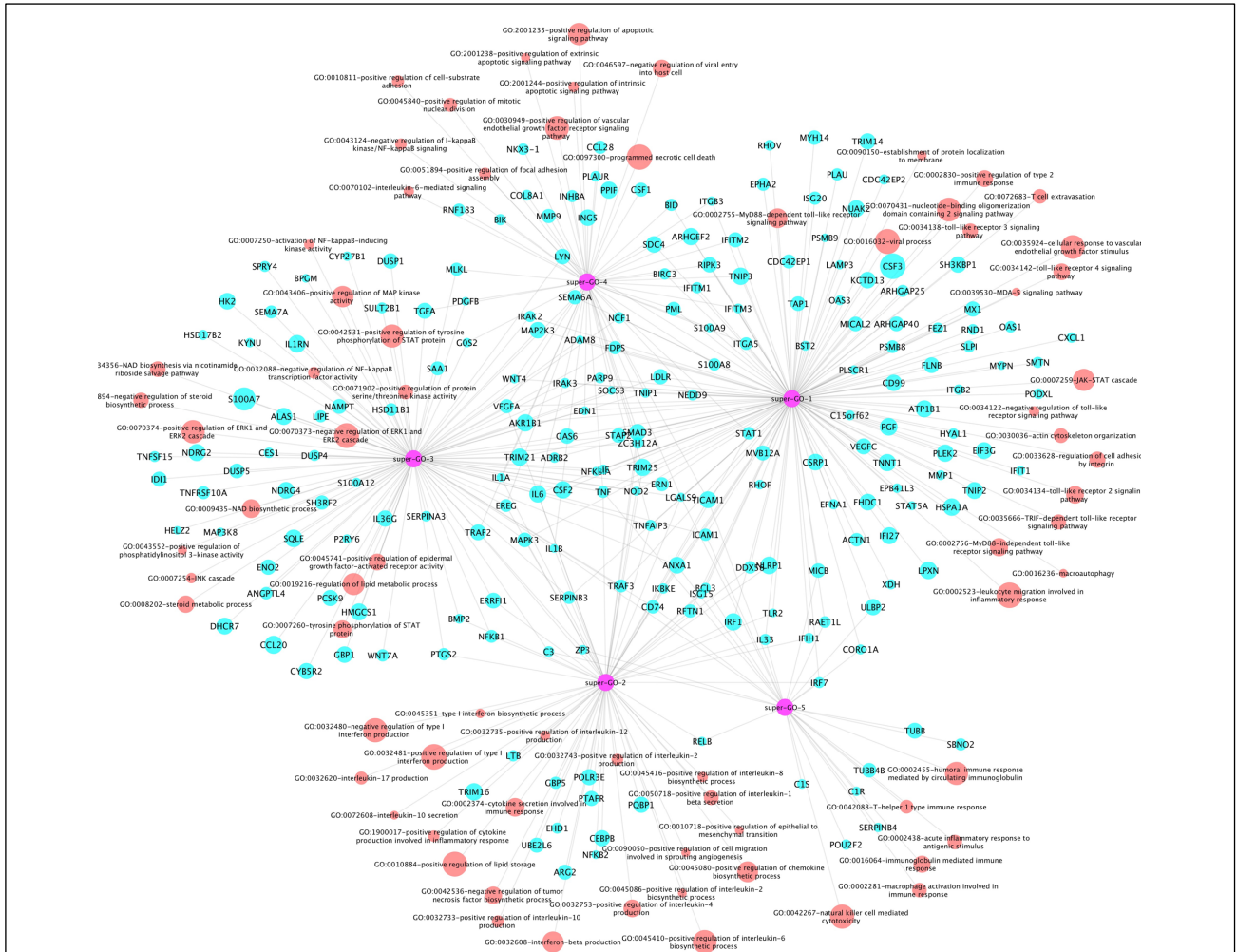

**Figure 2**: Network of 212 up-regulated genes (cyan color) in lung NHBE, A549_ACE2 and CALU-3 cells after SARS-CoV-2. There are 73 activated gene ontologies (GOs; red color), and 5 super-GOs (clusters of GOs; purple color). The node size of genes and GOs is proportional to the fold change and negative log2 p-values respectively.

*3.3 Repurposing drugs inhibiting individual super-GOs*

The gene signatures in each of 5 super-GO clusters were used as the query input of the CMAP database to identify drugs capable of inhibiting the associated genetic pathways. Specifically, 255 drugs appeared in the top 100 drugs in at least one of the 5 super-GOs. There were 26 common drugs among the total drugs derived from aforementioned signaling network analysis and the super-GO analysis (see **Table II**). Sixty-five drugs were selected based on their frequency (frequency >= 3) appeared in the top 100 drugs of each of the 5 super-GOs (see **Fig. 3** and **Table III)**. In **Table III**, some frequency values are greater than 5, as some drugs were tested multiple times in CMAP database. Surprisingly, the dexamethasone (glucocorticoid receptor agonist, corticosteroid agonist, immunosuppressant) (frequency = 17, ranked as No. 1 in our prediction, see **Table III**), an existing FDA approved drug, was recently reported to be the first drug that can significantly reduce the death rate of COVID-19 patients receiving respiratory support[5]. Specifically, the clinical trials results indicated that dexamethasone reduced death rate by one-third in patients receiving invasive mechanical ventilation, and reduced the death rate by one-fifth in COVID-19 patients receiving oxygen without invasive mechanical ventilation[5]. and hydrocortisone (corticosteroid agonist, glucocorticoid receptor agonist, immunosuppressant, interleukin receptor antagonist) (frequency = 9) were reported to be related to COVID-19 treatment[23].

**Table III**: Top 65 drugs frequently appeared in the top-ranked drugs of 5 super-GOs.

| Name | Freq | Name | Freq | Name | Freq | Name | Freq | Name | Freq |
| --- | --- | --- | --- | --- | --- | --- | --- | --- | --- |
| dexamethasone | 17 | azithromycin | 4 | norepinephrine | 4 | doconexent | 3 | nitrendipine | 3 |
| hydrocortisone | 9 | carbetocin | 4 | promazine | 4 | efavirenz | 3 | olanzapine | 3 |
| atorvastatin | 7 | deferiprone | 4 | rilmenidine | 4 | estrone | 3 | phensuximide | 3 |
| fenofibrate | 5 | diazepam | 4 | simvastatin | 4 | fluticasone | 3 | piperacillin | 3 |
| flupirtine | 5 | doxycycline | 4 | verapamil | 4 | formoterol | 3 | pirfenidone | 3 |
| palonosetron | 5 | enalapril | 4 | ziprasidone | 4 | fostamatinib | 3 | pirlindole | 3 |
| parthenolide | 5 | gefitinib | 4 | amcinonide | 3 | lapatinib | 3 | pravastatin | 3 |
| pindolol | 5 | halcinonide | 4 | amoxapine | 3 | lenalidomide | 3 | quinidine | 3 |
| sitagliptin | 5 | iloperidone | 4 | amylocaine | 3 | mepyramine | 3 | rosuvastatin | 3 |
| testosterone | 5 | lovastatin | 4 | budesonide | 3 | naftifine | 3 | scopolamine | 3 |
| triamcinolone | 5 | melperone | 4 | dicloxacillin | 3 | naloxone | 3 | temozolomide | 3 |
| trimethobenzamide | 5 | memantine | 4 | diethylstilbestrol | 3 | naltrexone | 3 | thalidomide | 3 |
| vemurafenib | 5 | mestranol | 4 | diltiazem | 3 | nifedipine | 3 | tocainide | 3 |

We further compared the GO analysis derived drugs with the clinical trials drugs, and 16 drugs were in both the prediction list and the clinical trial reports. The 16 drugs are: dexamethasone, lopinavir, chloroquine, leflunomide, doxycycline, sitagliptin, lenalidomide, simvastatin, verapamil, naproxen, tacrolimus, tofacitinib, pirfenidone, thalidomide, formoterol, and azithromycin. The lopinavir (HIV protease inhibitor) was also identified from top-ranked drugs against the super-GOs, which has been used for coronavirus treatment in clinical trials. Moreover, 6 drugs appeared in the predictions derived from signaling network analysis, GO analysis and clinical trials: chloroquine, lenalidomide, simvastatin, naproxen, tofacitinib and thalidomide. As seen, many of the predicted drugs were reported in the clinical trials or literature reports, which indicated that the prediction analysis could be helpful for repurposing existing drugs as novel and potentially effective treatments for COVID-19.

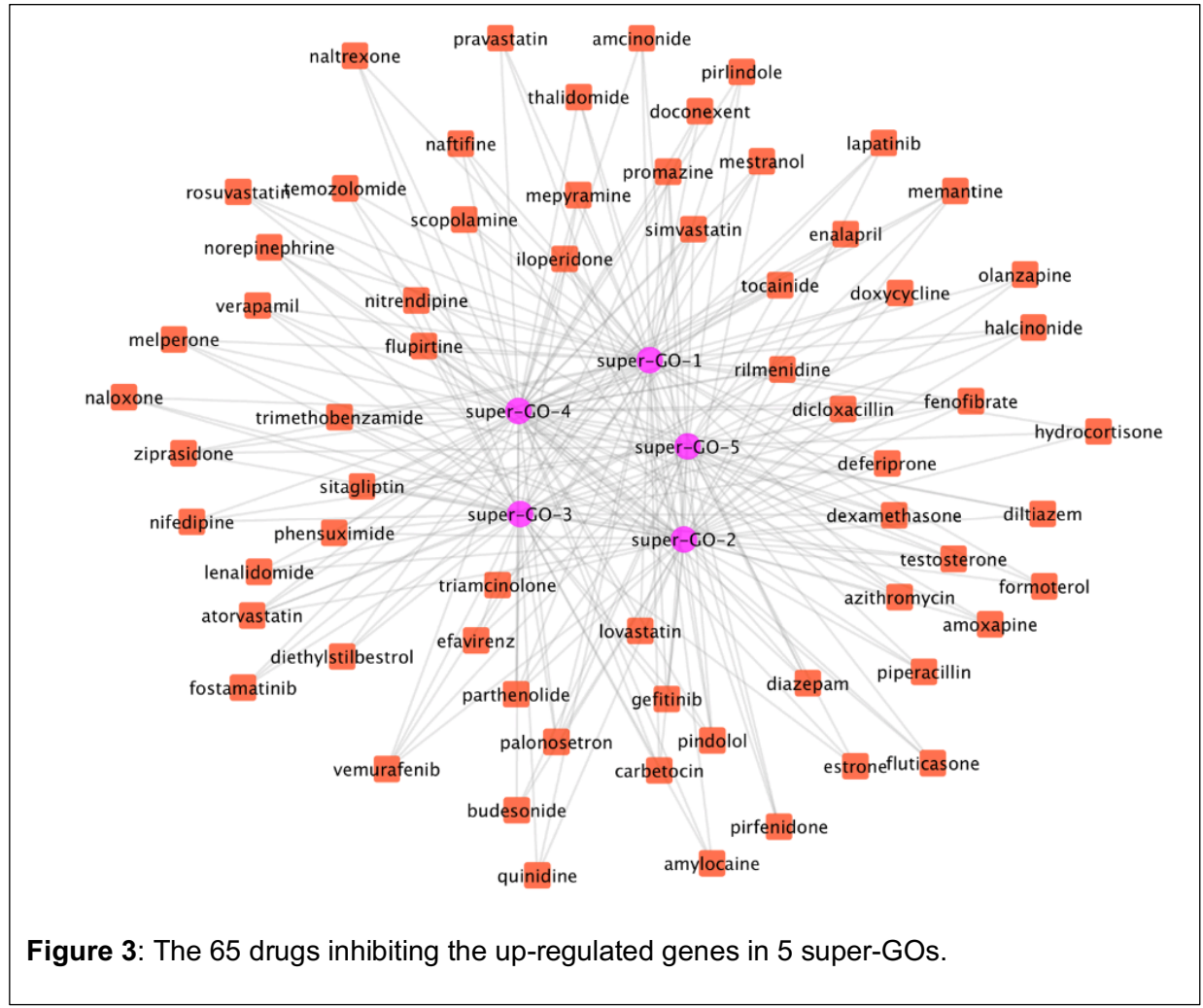

**Figure 3**: The 65 drugs inhibiting the up-regulated genes in 5 super-GOs.

## 4. Discussion and Conclusion

Currently, there is no effective new drugs or vaccine approved for the treatment of COVID-19, though rapid developments are occurring and being translated into clinical use. As the disease continues to spread, it is becoming increasingly important to develop a treatment modality in the most time-efficient manner. One such approach is to use genetic information to inform the repurposing of available medications. This preliminary and exploratory analysis uses transcriptional response (gene expression) profiles from human host cells before and after the infection with the SARS-CoV-2. In the analysis results, some potentially important targets, signaling pathways, and a set of GOs activated within host cells after viral infection, were identified. Moreover, a set of drugs registered for COVID-19 treatment globally were also identified in the analysis. These discoveries can be helpful to facilitate the design of future clinical trials for the COVID-19 treatment.

This exploratory computational study still has some limitations that can be further improved in the future work. First, the genetic data are derived from an *in vitro* analysis and will inevitably have a gene expression profile different from an *in vivo epithelial,* which may be further modified on a person-to-person basis. We will investigate the gene expression profiles from COVID-19 patients in the future work. Second, the signaling network analysis models could be improved by integrating the KEGG signaling pathways with the GOs (to include more genes) to uncover the core signaling networks involved in the life cycle of SARS-CoV-2 within host cells. In the future work, we will investigate these challenges. Third, the unbiased list of medications generated and presented was not filtered by route of administration or clinical applicability. For instance, the immune dampening chemotherapeutic agent docetaxel was identified, but would likely not be administered to an infected patient due to concern for augmenting viral replication. Therefore, further pipelines and additional information are needed to analyze the potential effects of these medications, and to continue developing this computational approach to medication repurposing.

**Acknowledgement**
The authors would like to thank Ms Kendall Cornick and Ms Kelley Foyil for proofreading the manuscript.

## Tables

**Table S1**: 220 drugs and targets on the potential dysfunctional signaling network of COVID19.

| Gene Symbol | Drug Name | Gene Symbol | Drug Name | Gene Symbol | Drug Name | Gene Symbol | Drug Name |
|---|---|---|---|---|---|---|---|
| BCL2L1 | isosorbide | IL1A | rilonacept | NCF1 | dextromethorphan | RELA | dimethyl fumarate |
| CASP1 | minocycline | IRAK4 | fostamatinib | NCF2 | dextromethorphan | RIPK1 | fostamatinib |
| CSNK2A2 | fostamatinib | ITGB2 | simvastatin | NFATC1 | pseudoephedrine | RIPK2 | fostamatinib |
| CTNNB1 | urea | LEF1 | etacrynic acid | NFKBIA | acetylsalicylic acid | RPS6KA4 | flavin mononucleotide |
| CXCL10 | clove oil | LTA | etanercept | NOD2 | mifamurtide | SGK1 | flavin mononucleotide |
| CXCL12 | tinzaparin | MAP2K3 | fostamatinib | NOX1 | oxygen | STAT5B | dasatinib |
| ERN1 | fostamatinib | MAP3K11 | fostamatinib | PAK1 | fostamatinib | SYK | fostamatinib |
| FLNA | artenimol | MAP3K13 | fostamatinib | PAK4 | fostamatinib | TAB1 | manganese |
| GJA1 | carvedilol | MAP4K4 | fostamatinib | PKN1 | fostamatinib | TBK1 | fostamatinib |
| HIF1A | carvedilol | MAPK11 | regorafenib | POMC | loperamide | TEC | fostamatinib |
| IFNGR1 | interferon gamma-1b | MAPK13 | fostamatinib | PTK2 | fostamatinib | TLR8 | imiquimod |
| IKBKE | fostamatinib | MYC | nadroparin | RAC2 | dextromethorphan | TNFRSF1A | tasonermin |
| ZAP70 | fostamatinib | CSNK2A1 | resveratrol, fostamatinib | FYN | dasatinib, fostamatinib | IL1R1 | anakinra, foreskin fibroblast (neonatal) |
| AKT1 | arsenic trioxide, resveratrol | FOS | pseudoephedrine, nadroparin | GSK3B | lithium, fostamatinib | IL6R | tocilizumab, sarilumab |
| ITK | pazopanib, fostamatinib | MAPK14 | dasatinib, fostamatinib | PIK3CA | caffeine, copanlisib | RAC1 | dextromethorphan, azathioprine |
| MAP3K1 | binimetinib, fostamatinib | PDPK1 | celecoxib, fostamatinib | PRKCI | tamoxifen, fostamatinib | ROCK1 | ripasudil, netarsudil |
| TLR2 | lyme disease vaccine (recombinant ospa), tuberculin purified protein derivative | TNFRSF1B | etanercept, tasonermin | MAP2K1 | cobimetinib, bosutinib, trametinib, | BTK | dasatinib, ibrutinib, acalabrutinib, fostamatinib |
| TLR7 | imiquimod, hydroxychloroquine | ICAM1 | natalizumab, hyaluronic acid, nafamostat | MAPK1 | isoprenaline, arsenic trioxide, turpentine, | CHUK | aminosalicylic acid, mesalazine, sulfasalazine, acetylcysteine |
| TLR9 | chloroquine, hydroxychloroquine | LY96 | morphine, methadone, lauric acid, | VCAM1 | ethanol, carvedilol, clove oil, | JAK1 | ruxolitinib, tofacitinib, baricitinib, fostamatinib |
| JAK2 | ruxolitinib, tofacitinib, baricitinib, fostamatinib | TLR4 | naloxone, lauric acid, papain, mifamurtide | ITGAL | efalizumab, antithymocyte immunoglobulin (rabbit), lovastatin, simvastatin, lifitegrast | IL1B | minocycline, gallium nitrate, canakinumab, rilonacept, foreskin keratinocyte (neonatal), binimetinib |
| JUN | vinblastine, pseudoephedrine, irbesartan, arsenic trioxide | EDNRA | bosentan, acetylsalicylic acid, sitaxentan, ambrisentan, macitentan, | ITGB3 | abciximab, eptifibatide, antithymocyte immunoglobulin (rabbit), tirofiban, resveratrol, | IL6 | ginseng, siltuximab, polaprezinc, foreskin fibroblast (neonatal), foreskin keratinocyte (neonatal), binimetinib |
| NFKB2 | acetylsalicylic acid, glucosamine, glycyrrhizic acid, fish oil | IFNG | olsalazine, glucosamine, apremilast, foreskin fibroblast (neonatal), foreskin keratinocyte (neonatal), | MTOR | pimecrolimus, sirolimus, everolimus, temsirolimus, fostamatinib, | NFKB1 | acetylsalicylic acid, thalidomide, pranlukast, triflusal, glycyrrhizic acid, fish oil |
| SRC | dasatinib, citric acid, bosutinib, ponatinib, nintedanib, fostamatinib | | | | | | |
| IKBKB | mesalazine, sulfasalazine, acetylsalicylic acid, auranofin, arsenic trioxide, acetylcysteine, fostamatinib, | | | | | | |
| PRKCA | phosphatidyl serine, vitamin e, tamoxifen, ingenol mebutate, midostaurin, alpha-tocopherol succinate, d-alpha-tocopherol acetate, | | | | | | |
| IFNAR1 | peginterferon alfa-2a, interferon alfa-n1, interferon alfa-n3, peginterferon alfa-2b, interferon alfa-2a, recombinant, interferon beta-1a, interferon beta-1b, interferon alfacon-1, interferon alfa-2b, recombinant, natural alpha interferon | | | | | | |
| KDR | sorafenib, sunitinib, ramucirumab, pazopanib, midostaurin, axitinib, cabozantinib, regorafenib, ponatinib, lenvatinib, nintedanib, fostamatinib | | | | | | |
| VEGFA | bevacizumab, minocycline, gliclazide, carvedilol, ranibizumab, pidolic acid, tromethamine, vandetanib, dalteparin, aflibercept, chondroitin sulfate, foreskin keratinocyte (neonatal) | | | | | | |
| EGFR | cetuximab, trastuzumab, lidocaine, gefitinib, erlotinib, lapatinib, panitumumab, vandetanib, afatinib, osimertinib, necitumumab, foreskin keratinocyte (neonatal), icotinib, neratinib, fostamatinib, brigatinib, olmutinib | | | | | | |
| TNF | etanercept, adalimumab, infliximab, chloroquine, epinephrine, pseudoephedrine, thalidomide, glucosamine, clenbuterol, pranlukast, amrinone, isopropyl alcohol, apremilast, golimumab, certolizumab pegol, pomalidomide, polaprezinc, foreskin fibroblast (neonatal), foreskin keratinocyte (neonatal), binimetinib, glycyrrhizic acid, | | | | | | |
| PTGS2 | dihomo-gamma-linolenic acid, icosapent, adapalene, aminosalicylic acid, mesalazine, acetaminophen, indomethacin, nabumetone, ketorolac, tenoxicam, lenalidomide, celecoxib, tolmetin, rofecoxib, piroxicam, fenoprofen, valdecoxib, diclofenac, sulindac, flurbiprofen, etodolac, mefenamic acid, naproxen, sulfasalazine, phenylbutazone, meloxicam, carprofen, diflunisal, suprofen, salicylic acid, meclofenamic acid, acetylsalicylic acid, bromfenac, oxaprozin, ketoprofen, balsalazide, thalidomide, ibuprofen, lumiracoxib, magnesium salicylate, salsalate, choline magnesium trisalicylate, ginseng, antrafenine, antipyrine, tiaprofenic acid, etoricoxib, resveratrol, niflumic acid, nimesulide, lornoxicam, aceclofenac, nepafenac, parecoxib, pomalidomide, loxoprofen, dexibuprofen, dexketoprofen, tolfenamic acid, morniflumate, propacetamol, talniflumate, phenyl salicylate, trolamine salicylate, menthyl salicylate, glycol salicylate, dipyrithione, alclofenac, bufexamac, acemetacin, fish oil, | | | | | | |